\documentclass[a4paper]{jpconf}
\usepackage{graphicx}
\usepackage{multirow}

\newcommand{\average}[1]{\ensuremath{\langle#1\rangle} }

\begin{document}
\title{Cold nuclear matter effect measured with high $p_{\rm T}$ hadrons and jets in 200GeV $d$+Au collisions in PHENIX}

\author{Takao Sakaguchi, for the PHENIX collaboration}

\address{Physics Department, Brookhaven National Laboratory, Upton, NY 11973, USA}

\ead{takao@bnl.gov}

\begin{abstract}
High $p_{\rm T}$ $\pi^0$ and $\eta$ as well as jets are measured using
high statistics $d$+Au collision data collected in RHIC Year-2008 run.
Both $R_{d{\rm A}}$ and $R_{\rm cp}$ for three observables are found to be
very consistent each other within quoted systematic and statistical
uncertainties. It was found that the $R_{d{\rm A}}$ is strongly centrality
dependent as opposed to the expectations from theoretical models.
An explanation of the centrality dependence from the experimental
point of view is presented.

\end{abstract}

\section{Introduction}
\begin{figure}[htbp]
\begin{minipage}{17pc}
\includegraphics[width=17pc]{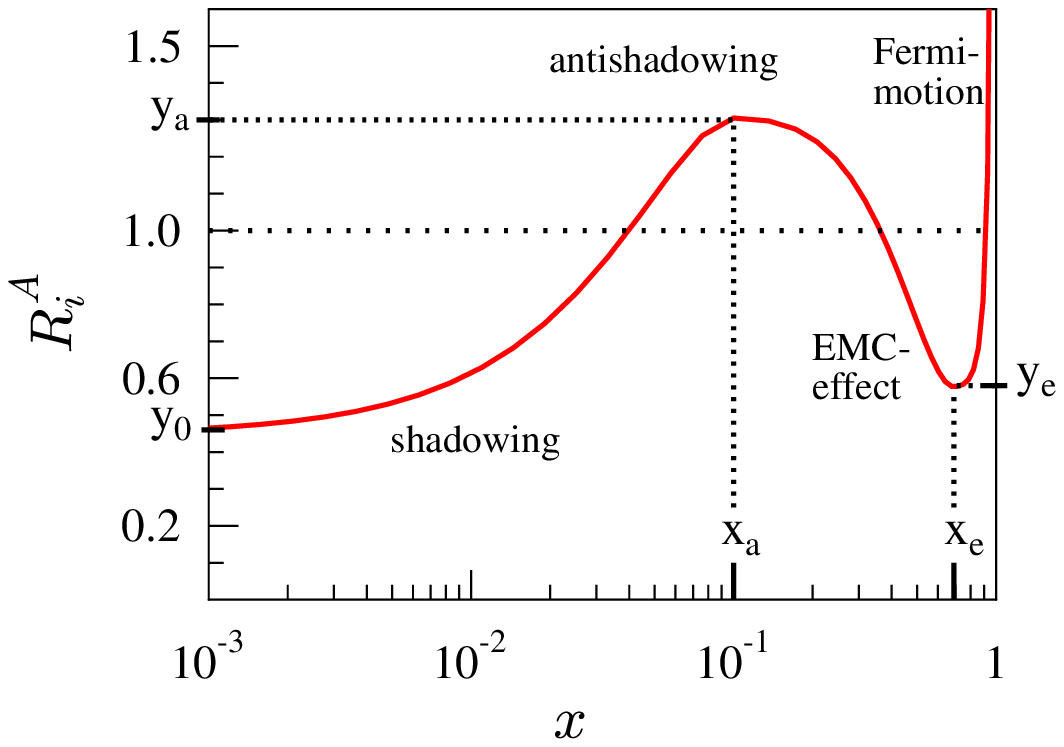}
\caption{\label{fig1_CNM} Nuclear modification ($R^{A}_{i}$(x)) to a free proton PDF in a nucleus.}
\end{minipage}\hspace{2pc}%
\begin{minipage}{19pc}
\includegraphics[width=19pc]{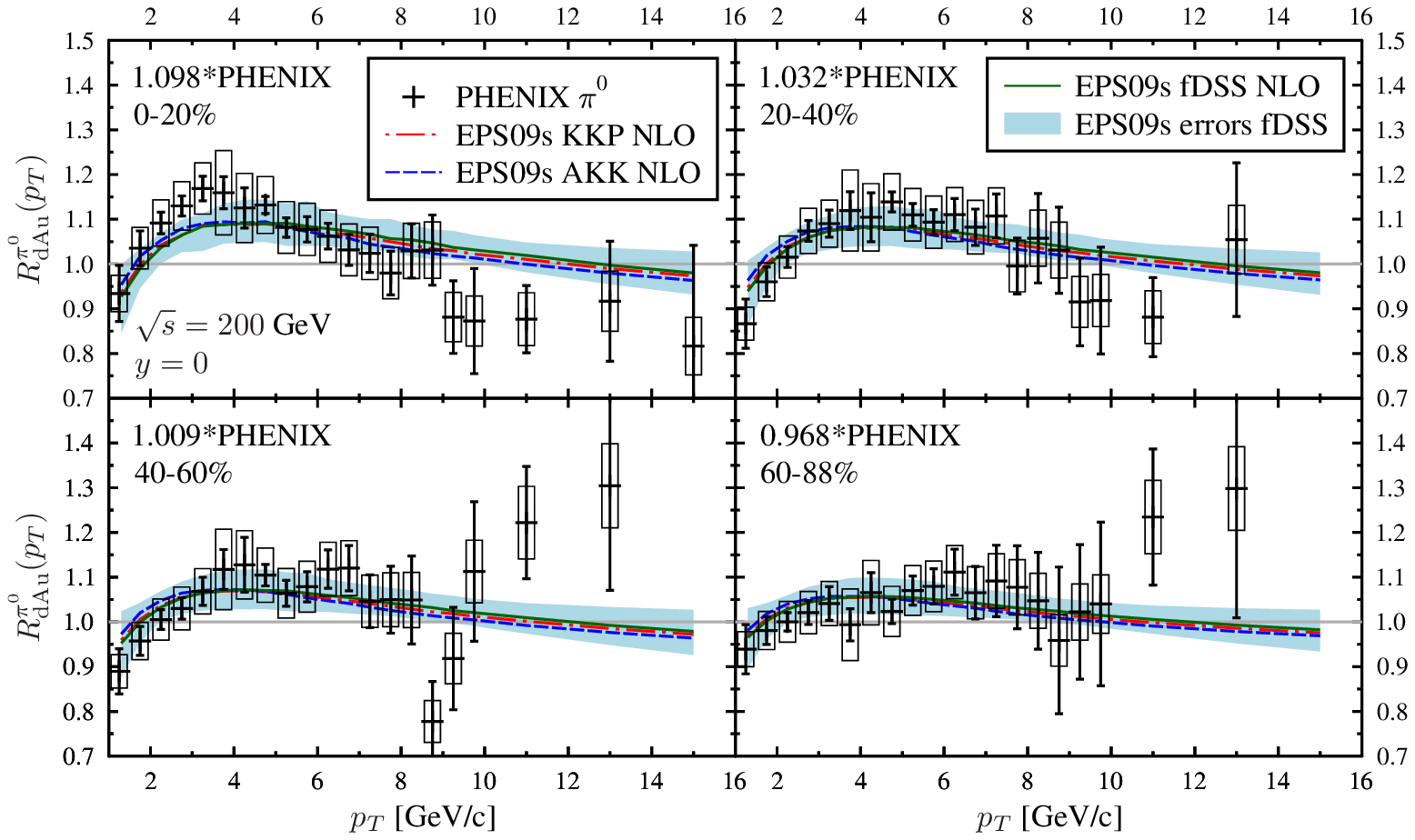}
\caption{\label{fig2_nPDF_Eskola}Comparison of measured $\pi^0$ $R_{d{\rm A}}$ and a theoretical model.}
\end{minipage} 
\end{figure}

It has been of interest that whether or not the initial hard scattering
process is modified in relativistic heavy ion collisions, after the
suppression of high transverse momentum ($p_{\rm T}$) hadron yields was
discovered in Au+Au collisions at $\sqrt{s_{NN}}$=130\,GeV in RHIC
Year-2000 run~\cite{ref1}. The first $d$+Au collision experiment was
carried out in
RHIC Year-2003 run and confirmed that the suppression attributes to a
final state interaction of hard scattered partons with a medium created
in Au+Au collisions~\cite{ref2}. After these results were published,
many theoretical works have been dedicated to
precisely determine the initial state effect that affects to the evaluation
of the final state effect such as parton energy loss. Figure~\ref{fig1_CNM}
shows a nuclear modification ($R^{A}_{i}$(x)) to a free proton PDF from the
CTEQ6.1M set~\cite{ref3} in the $\overline{MS}$ scheme, where $i$ denotes
the parton flavor~\cite{ref4}.
The hard scattering process is factorized into a fragmentation function,
a reaction cross-section, and a parton distribution function (PDF). Since
the $d$+Au collisions will not produce a bulk hot dense medium, the yield
difference of the hard scattering probe in $d$+Au and $p+p$ collisions are
predominantly from the modification of PDF. Therefore, the nuclear
modification factor
$R_{d{\rm A}}$ ($\equiv(dN_{d{\rm A}}/dydp_{T})/(\average{T_{d{\rm A}}}d\sigma_{pp}/dydp_{T})$)
is expected to reflect the modification of PDF. Eskola and companies tried
to construct the centrality dependent nuclear PDF using the previously
published $\pi^0$ $R_{d{\rm A}}$ result as shown in
Figure~\ref{fig2_nPDF_Eskola}~\cite{ref4}.
Obviously, the statistical uncertainty is too large to observe the centrality
dependence of the nuclear PDF. We analyzed another $d$+Au collision data set
collected in RHIC Year-2008 in order to tackle this interesting subject.

\section{Measurements of $\pi^0$, $\eta$, and jets in PHENIX}
\begin{figure}[htbp]
\begin{minipage}{18pc}
\includegraphics[width=18pc]{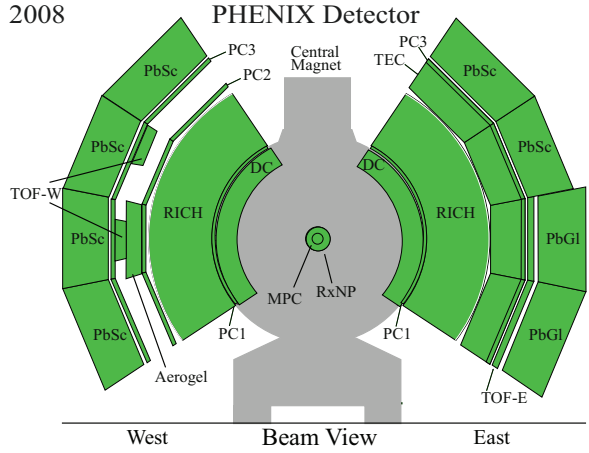}
\end{minipage}\hspace{2pc}%
\begin{minipage}{18pc}
\includegraphics[width=18pc]{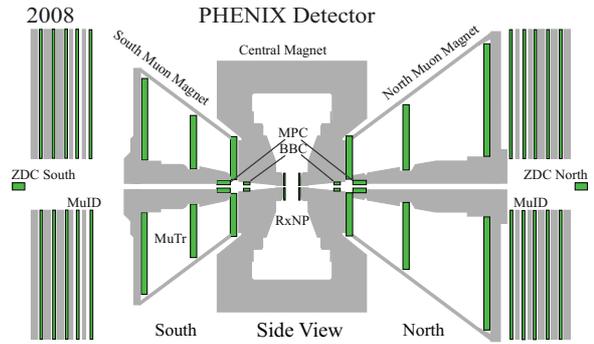}
\end{minipage} 
\caption{\label{fig3_PHENIX_run8}PHENIX detector in RHIC Year-2008 run.}
\end{figure}

\begin{figure}[htbp]
\includegraphics[width=38pc]{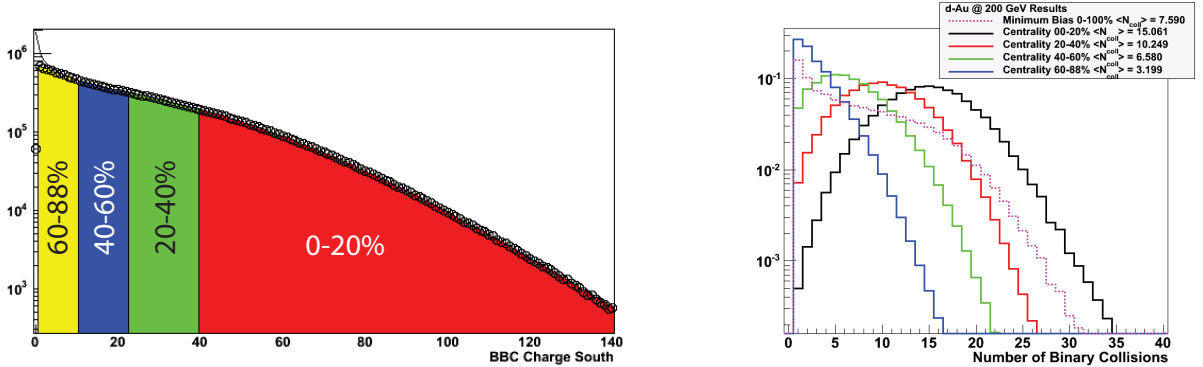}
\caption{\label{fig4_bbc}BBC south charge (left) and $N_{coll}$ (right) distributions for each centrality obtained from decomposition of the BBC charge distribution.}
\end{figure}

The schematic view of the PHENIX detector~\cite{ref5} in RHIC Year-2008 is
shown in Figure~\ref{fig3_PHENIX_run8}.
PHENIX sampled an integrated luminosity of 80\,nb$^{-1}$ during that run,
a factor of 30 increase over
the RHIC Year-2003 $d$+Au data set. The minimum bias events are triggered
by a coincidence of the signals from the beam-beam counters (BBC) located
at the south and north side of the detector system ($3.1<|\eta|<3.9$).
We define the forward (positive) rapidity as the Au-going side (pointing
to the south BBC). The collision centrality is determined using the
charge signal from the south BBC.
The underlying average number of binary nucleon-nucleon collisions
($N_{coll}$) and the number of participant nucleons ($N_{part}$) are
estimated with a Glauber model Monte Carlo simulation. Then
the result is combined with a negative binomial distribution (NBD)
with parameters ($\mu$, $k$) to model the charge distribution from
the south BBC. Figure~\ref{fig4_bbc} shows the charge distribution
in the south BBC (left) and the decomposed $N_{coll}$ distributions for
each centrality (right), respectively.
Due to the auto-correlation between having a high $p_{\rm T}$ particle
in the central PHENIX spectrometer and the charge multiplicity in the
BBC, an additional correction factor is calculated. This is done with
the same Glauber Monte Carlo simulation used for the calculation of
$N_{coll}$.

$\pi^0$'s and $\eta$'s are reconstructed through 2$\gamma$'s measured
with eight electromagnetic calorimeters (EMCal)~\cite{ref6, ref7}.
The EMCal are located most outside of the central arm detector system,
which covers an acceptance of $|\eta|<0.35$ and a half azimuth. The
contribution of random $\gamma\gamma$ combination is estimated using the
mixed event technique, and subtracted from the measured foreground
distribution. The drift chamber (DC) and pad chamber (PC) are also
used to veto charged particles including charged hadrons and electrons.
The jets are reconstructed using neutral electromagnetic clusters
measured in EMCal and charged particles measured in DC and PC.
Those clusters/particles with reconstructed $p_{\rm T}$ of
$p_{\rm T}>400$\,MeV/$c$ are included in the jet reconstruction procedure.
We employed Gaussian filtering algorithm with a filter width of
$\sigma$=0.3. This algorithm was also used in $p+p$ and Cu+Cu collision
analysis at PHENIX, and cross-checked with anti-$k_{\rm T}$ algorithm.
\begin{table}[htbp]
\caption{\label{tab1_syserr} Systematic uncertainties for $\pi^0$, $\eta$, and jet $p_{\rm T}$ measurement.}
\begin{center}
\begin{tabular}{c|c||c||c|c|c|c}
\hline\hline
\multirow{7}{5mm}{$\pi^0$}
 & source & type & 5\,GeV/$c$ & 10\,GeV/$c$ & 15\,GeV/$c$ & 20\,GeV/$c$ \\
\cline{2-7}\cline{2-7}
& Peak extraction & B & 2\,\% & 2\,\% & 2\,\% & 2\,\% \\
& Acceptance & C & 2.5\,\% & 2.5\,\% & 2.5\,\% & 2.5\,\% \\
& PID efficiency & B & 7\,\% & 8\,\% & 8.5\,\% & 9\,\% \\
&Energy scale & B & 7.5\,\% & 8\,\% & 8\,\% & 8\,\% \\
&Photon conversion & C & 2\,\% & 2\,\% & 2\,\% & 2\,\% \\
&Cluster merging & B & 0\,\% & 0\,\% & 8\,\% & 18\,\% \\
\hline\hline
\multirow{6}{5mm}{$\eta$}
 & source & type & 5\,GeV/$c$ & 10\,GeV/$c$ & 15\,GeV/$c$ & 20\,GeV/$c$ \\
\cline{2-7}\cline{2-7}
& Peak extraction & B & 2\,\% & 2\,\% & 2\,\% & - \\
&Acceptance & C & 2.5\,\% & 2.5\,\% & 2.5\,\% & - \\
&PID efficiency & B & 7\,\% & 8\,\% & 8.5\,\% & - \\
&Energy scale & B & 11\,\% & 12\,\% & 12\,\% & - \\
&Photon conversion & C & 2\,\% & 2\,\% & 2\,\% & - \\
\hline\hline
\multirow{6}{5mm}{jets}
 & source & type & 9-12\,GeV/$c$ & 12-15\,GeV/$c$ & 15-20\,GeV/$c$ & 20+\,GeV/$c$\\
\cline{2-7}\cline{2-7}
& Trigger efficiency & B & 5-8\,\% & 5-8\,\% & 5\,\% & 5\,\% \\
&Unfolding & B & 4-10\,\% & 4-10\,\% & 4-6\,\% & 4-6\,\% \\
&Residual fake rate & B & 5\,\% & 5\,\% & 5\,\% & 5\,\% \\
&Acceptance & C & 3\,\% & 3\,\% & 3\,\% & 3\,\% \\
&Vertex dependence & C & 1\,\% & 1\,\% & 1\,\% & 1\,\% \\
\hline\hline
\end{tabular}
\end{center}
\end{table}
A mild underlying events in $d$+Au collisions shifts the measured $p_{T}$
to a higher value than the true one. This effect is evaluated and
corrected for by embedding jets into real events. A small residual
fake rate ($<$5\,\%) was observed for $p_{\rm T}>$9\,GeV/$c$ and was
accounted in a systematic uncertainty.
The systematic uncertainties for $\pi^0$, $\eta$, and jets $p_{T}$
measurements are summarized in Table~\ref{tab1_syserr}.
The type-B errors are the errors correlated from $p_T$ to $p_T$. The type-C
errors are the errors of overall normalizations.

\section{Results}
\begin{figure}[htbp]
\begin{minipage}{18pc}
\includegraphics[width=18pc]{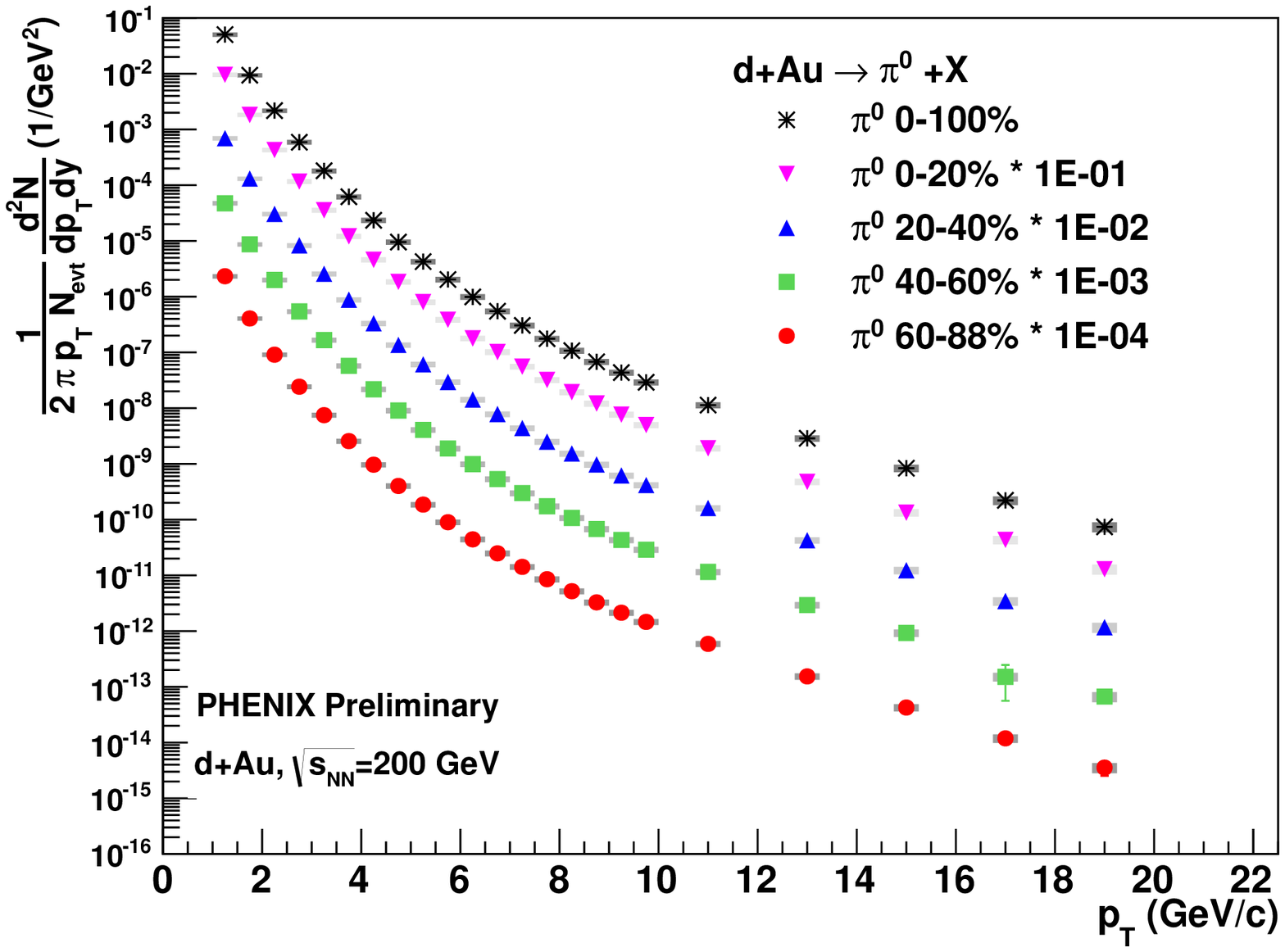}
\caption{\label{fig5_pi0_spectra}$p_{\rm T}$ spectra of $\pi^0$ measured in $d$+Au collisions.}
\end{minipage} 
\begin{minipage}{18pc}
\includegraphics[width=18pc]{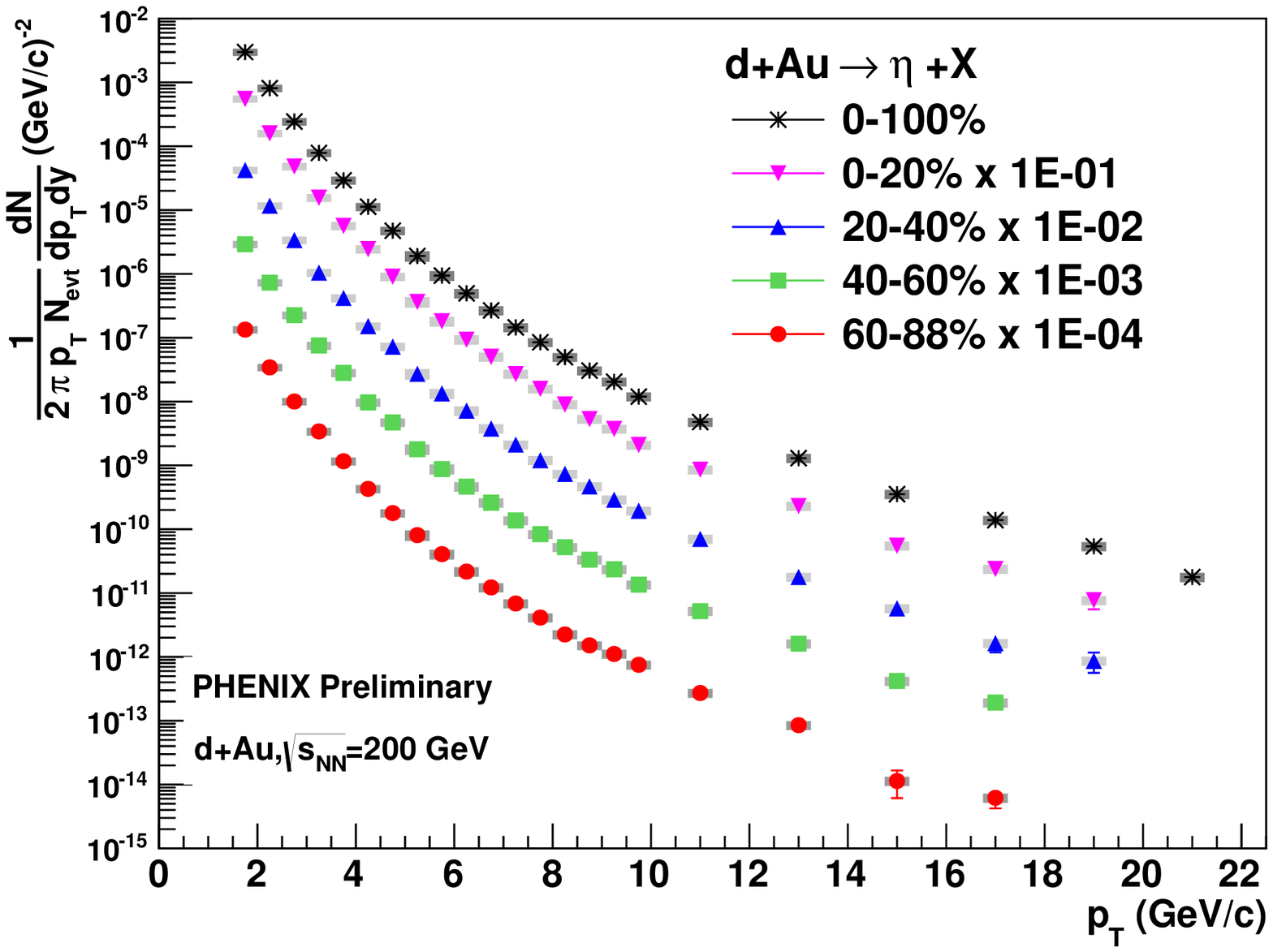}
\caption{\label{fig6_eta_spectra}$p_{\rm T}$ spectra of $\eta$ measured in $d$+Au collisions.}
\end{minipage} 
\end{figure}

\begin{figure}[htbp]
\begin{minipage}{20pc}
\includegraphics[width=20pc]{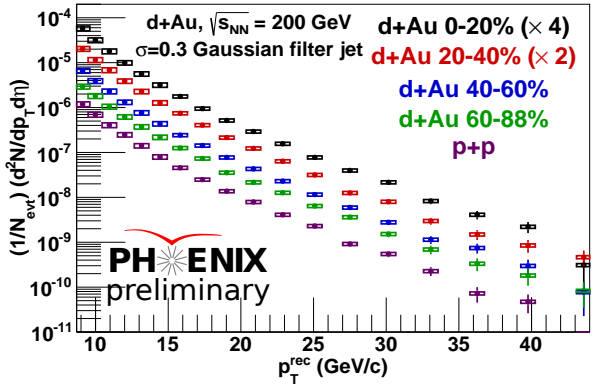}\hspace{2pc}%
\end{minipage}
\begin{minipage}[t]{16pc}
\vspace{-8mm}
\caption{\label{fig7_jet_spectra}$p_{\rm T}$ spectra of jets measured in $d$+Au collisions.}
\end{minipage}
\end{figure}

\begin{figure}[htbp]
\begin{center}
\includegraphics[width=27pc]{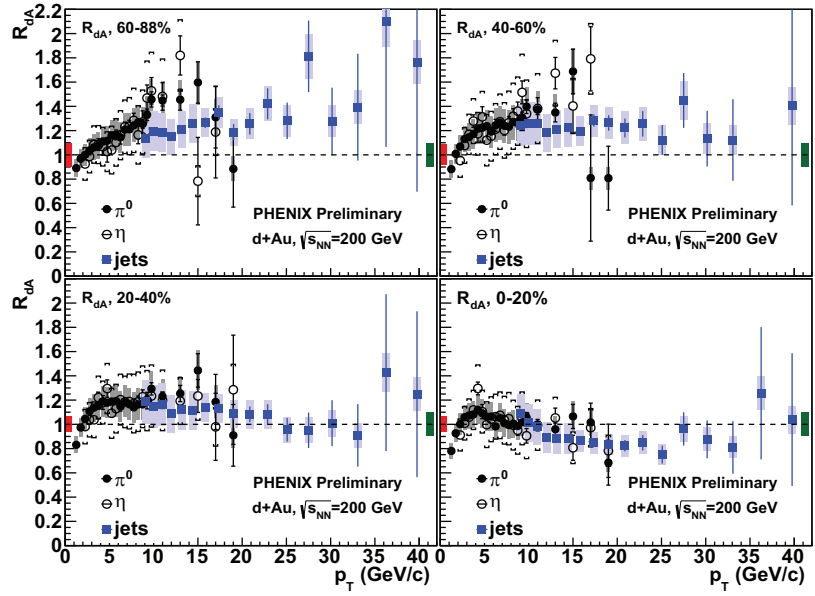}
\caption{\label{fig8_RdA}$R_{d{\rm A}}$ for $\pi^0$, $\eta$ and jets in $d$+Au collisions for various centralities.}
\end{center}
\end{figure}

The $p_{\rm T}$ spectra of $\pi^0$ and $\eta$ are shown in
Figures~\ref{fig5_pi0_spectra} and ~\ref{fig6_eta_spectra}, respectively.
With the better statistics in RHIC Year-2008 run, we extended $p_{\rm T}$
reach by more than 5\,GeV/$c$. As going to higher
$p_{\rm T}$ ($p_{\rm T}>\sim12$\,GeV/$c$), the opening angle of
$\pi^0\rightarrow\gamma\gamma$ becomes so small that they can not be
resolved in EMCal because of its limited granularity. It is called as the
cluster merging effect. An $\eta$ has a four times larger opening angle
($\theta_{\gamma\gamma}\propto M/p_{\rm T}$) and thus the merging of two
$\gamma$'s does not happen until $\sim$50\,GeV/$c$.
Due to this effect, the ratio of raw $\eta$ to $\pi^0$ yield increases
at higher $p_{\rm T}$~\cite{ref7}.
The $p_{\rm T}$ spectra of jets is shown in Figure~\ref{fig7_jet_spectra}.
Jets are reconstructed from multi-particles, resulting in reaching to much
higher $p_{\rm T}$, as high as 45\,GeV/$c$, with the same integrated
luminosity. We computed the $R_{d{\rm A}}$ for $\pi^0$, $\eta$, and jets as
shown in Figure~\ref{fig8_RdA}.
\begin{figure}[htbp]
\begin{center}
\includegraphics[width=29pc]{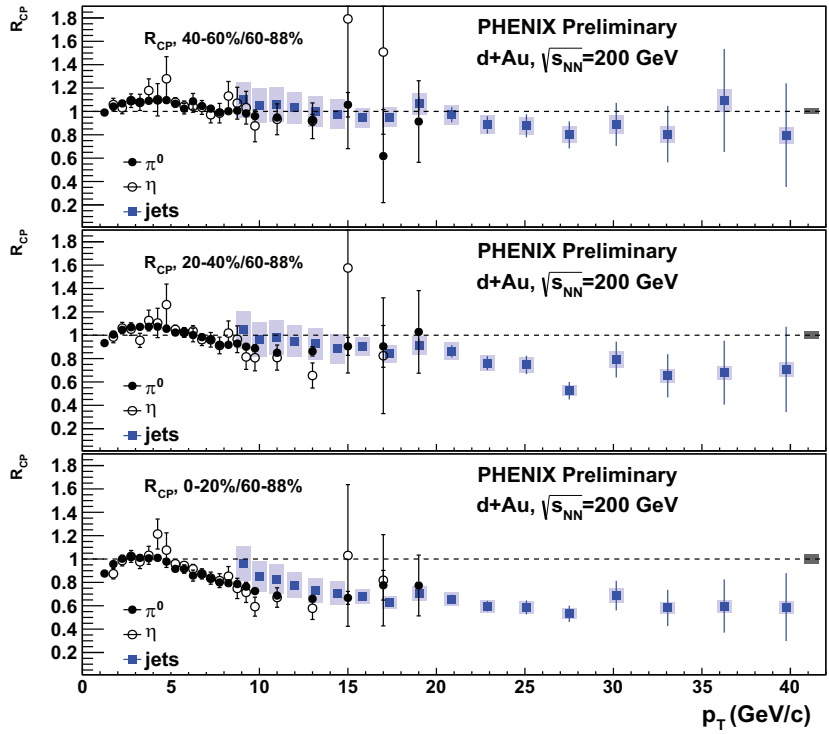}
\caption{\label{fig9_Rcp}$R_{\rm cp}$ for $\pi^0$, $\eta$ and jets in $d$+Au collisions for various centralities. The 60-88\,\% centrality is taken as the denominator.}
\end{center}
\end{figure}
Although jets and single identified hadrons can not be
directly compared at the same $p_{\rm T}$, they are agreeing each other
very well within quoted uncertainties. For reader's sake, it may be
useful to note that the mean $z$($\equiv p_{\rm T}^{had}/p_{\rm T}^{jet}$)
for $\pi^0$ and $\eta$ is $\sim$0.7.
It was found that the $R_{d{\rm A}}$ is strongly centrality dependent; the most
central collisions show a small suppression, while the most peripheral
collisions show a sizable enhancement, compared to the expectation from
$p+p$ collisions. It should also be noted that the enhancement in the
peripheral collisions increases as going to higher $p_{\rm T}$.
This trend shows that the nuclear PDF may be strongly dependent of impact
parameters as opposed to what was predicted by theoretical models.
In order to investigate the suppression/enhancement trend with reduced
systematic uncertainties, the ratio of the yield in central to peripheral
collisions ($R_{\rm cp}$) is computed and shown in Figure~\ref{fig9_Rcp}.
In this ratio, the error from $p+p$ baseline measurement will be excluded.
It was found that the $\pi^0$, $\eta$, and jet $R_{\rm cp}$ show the same
trend and are consistent each other within even smaller uncertainties.
The decreasing trend as going to higher $p_{\rm T}$ is consistent with the
fact that the $R_{d{\rm A}}$ is enhanced in peripheral collisions and
suppressed in central collisions.

\section{Global feature in $d$+Au collisions}
\begin{figure}[htbp]
\begin{center}
\begin{minipage}{28pc}
\includegraphics[width=28pc]{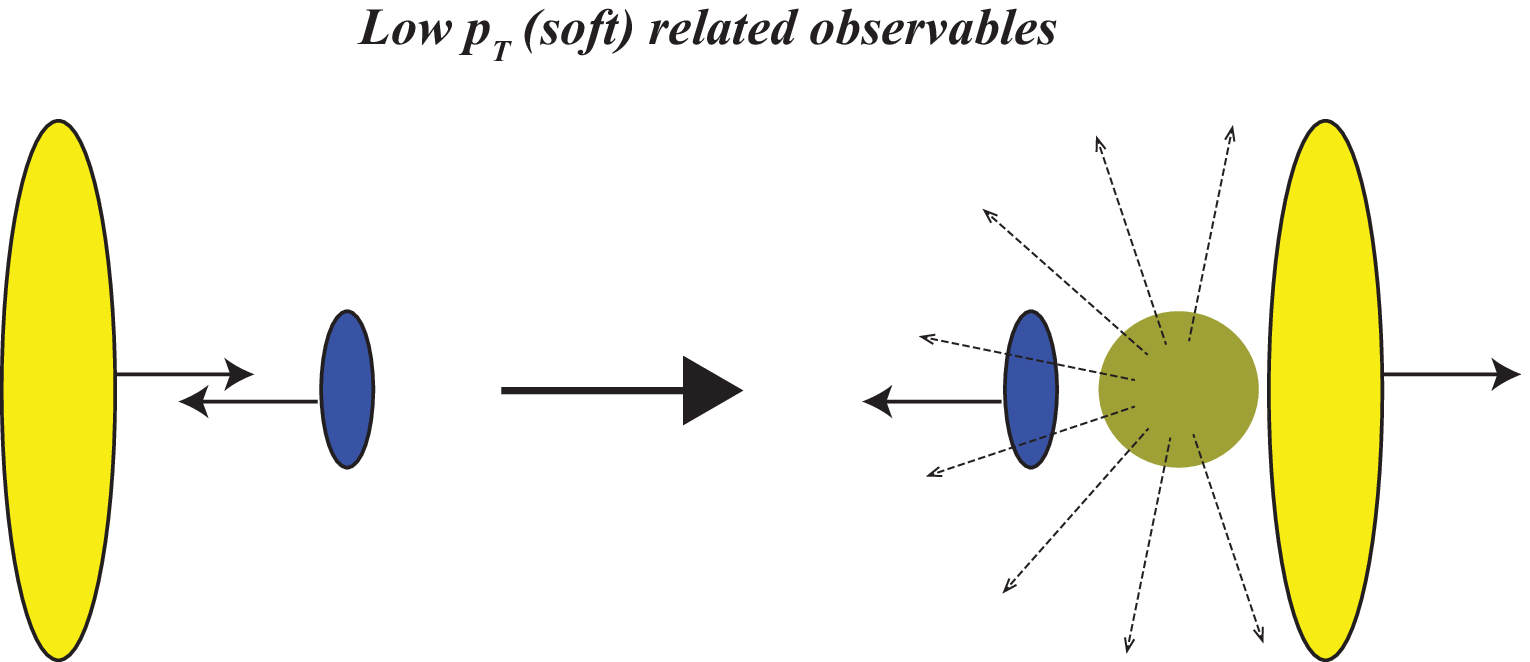}
\end{minipage} \\
\vspace{5mm}
\begin{minipage}{28pc}
\includegraphics[width=28pc]{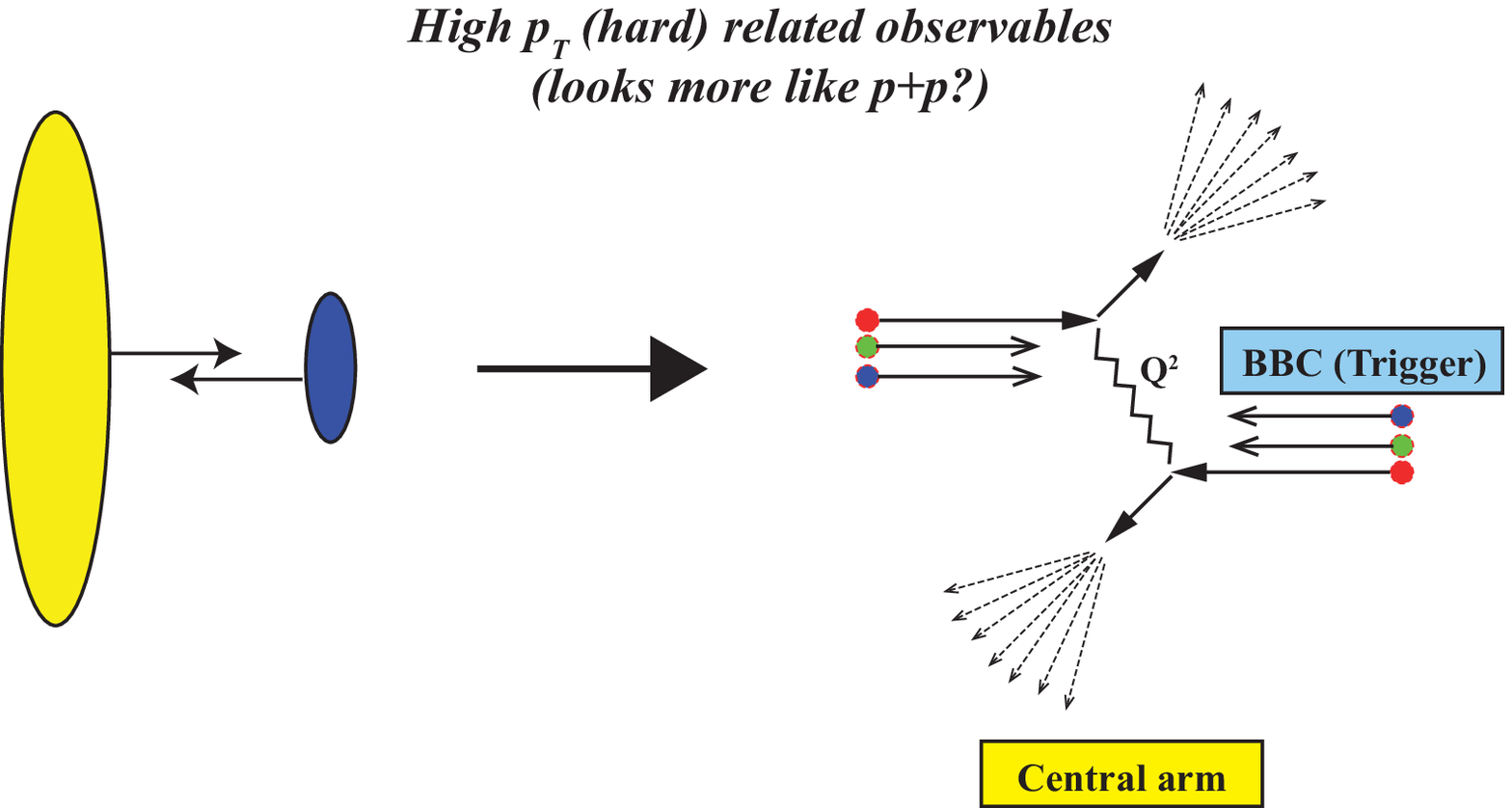}
\end{minipage} 
\caption{\label{fig10_hardfact}Collision dynamics for events with low $p_{\rm T}$ (top) and high $p_{\rm T}$ (bottom) observables.}
\end{center}
\end{figure}

\begin{figure}[htbp]
\begin{center}
\includegraphics[width=30pc]{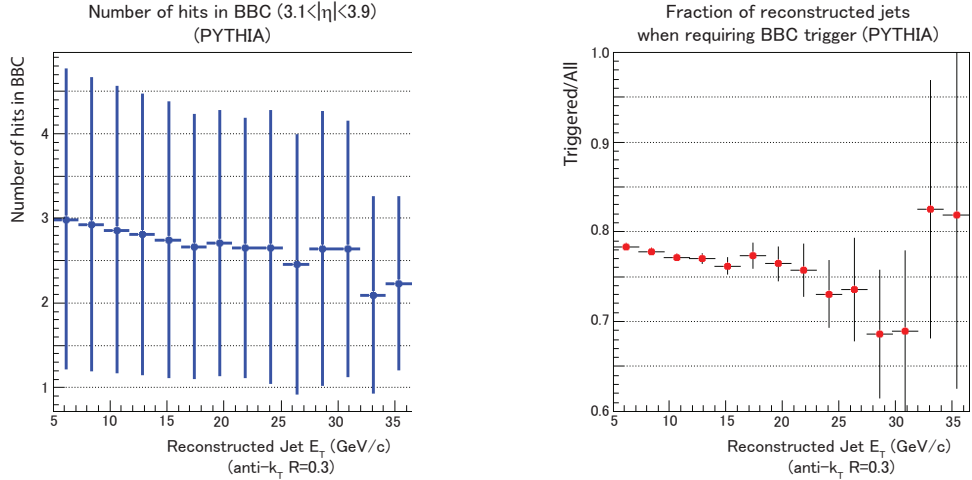}
\caption{\label{fig11_ppbias_PYTHIA}PYTHIA simulation of the number of BBC hits (left) and the fraction of jets seen in central arm when requiring BBC hits (right).}
\end{center}
\end{figure}

\begin{figure}[htbp]
\begin{center}
\includegraphics[width=30pc]{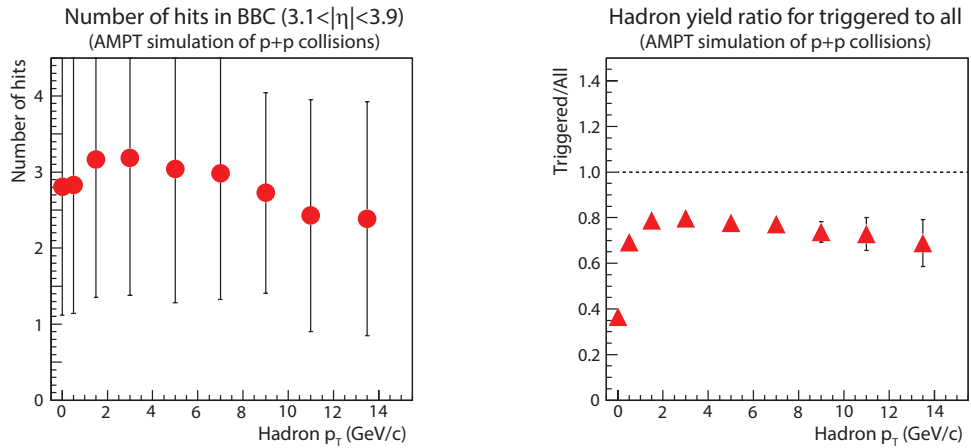}
\caption{\label{fig12_ppbias_AMPT}AMPT simulation of the number of BBC hits (left) and the fraction of $\pi^0$'s seen in central arm when requiring BBC hits (right).}
\end{center}
\end{figure}

Before one concludes that the nuclear PDF is strongly impact parameter
dependent, one may want to revisit some global features in $d$+Au collisions
such as triggering or centrality definition.
In any collisions, one has to generate a trigger using signals
from detectors whose efficiencies are not 100\,\% in reality. The
efficiency of a trigger is often particle multiplicity dependent.
Therefore, when triggering minimum bias events in small collision
system such like $d$+Au, the peripheral collisions are most biased.
In Figure~\ref{fig10_hardfact}, we tried to sketch the collision
dynamics for events with low and high $p_{\rm T}$ observables
in peripheral collisions.
As explained before, we trigger minimum bias events and define their
centrality classes based on the charge signals from the south side
BBC (Au-going direction). As seen in Figure~\ref{fig10_hardfact},
the soft-process-dominated events emit particles isotropically.
Therefore, the BBC charge is proportional to the centrality.
However, in the case that hardest jets are produced, less energy
will be available for soft production at high $\eta$, e.g., at the
BBC location. This leads to a possible $p_{\rm T}$ dependence of the
trigger efficiency in most peripheral collisions.
We performed a simulation study for the simplest case, $p+p$ collisions,
using PYTHIA~\cite{ref8} and AMPT~\cite{ref9} event generators to see
this effect.
The left side of the Figure~\ref{fig11_ppbias_PYTHIA} shows the number
of BBC hits as a function of jet $E_{\rm T}$ obtained from a PYTHIA
simulation.
Since the collision system is symmetric, we took the BBC hit distributions
from both south and north sides. The error bars show RMS of the
distributions. As going to higher $E_{\rm T}$, the number of hits in BBC
decreases, and the trigger efficiency of events is reduced as shown in the
right side of the same figure. At $\sim$30\,GeV/c the
trigger efficiency becomes 70\,\%. Figure~\ref{fig12_ppbias_AMPT} shows
a AMPT simulation of $p+p$ minimum bias collisions.
In this figure, the number of BBC hits and the trigger efficiency are plotted
against the $p_{\rm T}$ of $\pi^0$'s. The yield of $\pi^0$'s starts
decreasing at $\sim$4\,GeV/$c$, and reaches to $\sim$30\,\% deficit at
14\,GeV/$c$. The trend and magnitude are found to be consistent with what
PYTHIA simulation showed, by taking the difference of the energy scale between
jets and $\pi^0$'s into account. This implies a trigger bias is expected
from a simple hard scattering kinematics, and manifests both in jet
$E_{\rm T}$ and single hadron $p_{\rm T}$.
We are now looking for a same effect in the $d$+Au collision case using
the simulations.

\section{Summary and Outlook}
High $p_{\rm T}$ $\pi^0$ and $\eta$ as well as jets are measured using high
statistics $d$+Au collision data collected in RHIC Year-2008 run. Both
$R_{d{\rm A}}$ and $R_{\rm cp}$ for three observables are found to be very
consistent each other within quoted systematic and statistical
uncertainties. It was found that the $R_{d{\rm A}}$ is strongly centrality
dependent as opposed to the expectations from theoretical models.
A possible explanation is given
based on the kinematics of hard scattering process. A simulation study
in $p+p$ collision case using PYTHIA and AMPT event generators is
presented for supporting the explanation. The study for $d$+Au collisions
is on-going.

Recently, a dedicated workshop on $d$+A and $p$+A collisions at RHIC and
LHC was organized~\cite{ref10}.
In the workshop, it was pointed out that the collision dynamics in $d$+Au
collisions is a sequence of 2$\rightarrow$2 hard processes, therefore,
the statistical treatment of collisions, such like using $N_{coll}$ may
lead a problem. We expect a rich physics in this small collision system.

\end{document}